\documentclass[
twocolumn
]{jpsj3}
\usepackage{multirow}
\makeatletter

\@addtoreset{equation}{section}
\makeatother
\title{Effect of Anharmonicity on the Kondo Phenomena of a Magnetic Ion Vibrating in a Confinement Potential}
\author{
Satoshi YASHIKI and Kazuo UEDA\thanks{E-mail address: ueda@issp.u-tokyo.ac.jp}
}
\inst{Institute for Solid State Physics, University of Tokyo, Kashiwa 277-8581, Japan}
\abst{Effect of anharmonicity of a cage potential for a magnetic ion vibrating in a metal is investigated by the 
numerical renormalization group method. The cage potential is assumed to be one-dimensional and of the double-well 
type. In the absence of the Coulomb interaction, we find continuous crossover among the three limiting cases: 
Yu-Anderson-type Kondo regime, the double-well-type Kondo one, and the renormalized Fermi chain one.  In the entire parameter 
space of the double-well potential, the ground state is described by a local Fermi liquid. In the Yu-Anderson-type 
Kondo regime, a quantum phase transition to the ground state with odd parity takes place passing through the two-channel 
Kondo fixed point when the Coulomb interaction increases. Therefore, the vibration of a magnetic ion in an oversized cage
structure is a promising route to the two-channel Kondo effect.}
\kword{Kondo effect, impurity Anderson model, anharmonic potential, electron-vibration coupling, numerical 
       renormalization group}
\begin{document}
\maketitle
\section{Introduction}
Recently, many researchers in condensed matter physics have been interested in characteristic ionic structures, 
networks of cages filled or unfilled by guest ions. One typical example is the filled skutterudite compounds 
RT$_{4}$X$_{12}($R $=$ rare earth or alkaline earth; T $=$ Fe, Ru, Pt or Os; X $=$ P, As, Ge or Sb$)$. In a certain case, 
the radius of filled ion is smaller than the size of cage. In such a case, it is expected that they will show various unusual 
physical behaviors due to the vibrations of the guest ions in strongly anharmonic potential.

A very peculiar feature is observed in SmOs$_{4}$Sb$_{12}$\cite{Sm_Os4_Sb12_1,Sm_Os4_Sb12_2,Sm_Os4_Sb12_3,Sm_Os4_Sb12_4,
Sm_Os4_Sb12_5}. It is reported that a large specific heat coefficient $\gamma$ is obtained, where the unusual phenomenon is 
its robustness against magnetic field\cite{Sm_Os4_Sb12_2}. Some theoretical studies propose as a possible scenario that local 
vibrations of the guest ions lead to the nonmagnetic Kondo effect\cite{Hattori,Hotta1,Hotta2}. In these theories, the authors consider
the situation where the guest ion moves back and forth among several potential minima in the cage potential. Actually, off-center 
modes of the guest ion are reported in several cage materials\cite{off_center1,off_center2,off_center3}. From 
a theoretical standpoint, there are two different types of local vibrations which couple with conduction electrons; the 
breathing mode of the cage and the transverse mode, namely relative displacement between the guest ion and the cage. 
We have studied the interplay between the electron correlation and the transverse type vibrations in Refs. 
\ref{yashiki1} and \ref{yashiki2}. However, in the previous studies, the cage potential was assumed to be harmonic.

Let us review former studies of the nonmagnetic Kondo effect concerning the transverse mode. Vlad$\acute{\text{a}}$r 
and Zawadowski considered the model where a single atom tunnels between two positions, called as the two level system (TLS).
They showed that the logarithmic divergence appears when the doubly degenerate states are connected each other through 
scattering processes of spinless conduction electrons\cite{TLM1,TLM2,TLM3}. This type of Kondo effect will be referred to 
as the double-well-type Kondo (DWK) in this paper. Subsequently, Yu and Anderson considered the first-order term of ionic 
displacement which produced the scattering processes between the spinless $s$-wave conduction electrons to the $p$-wave 
ones\cite{Yu_Anderson}. When the electron-phonon coupling is strong, the ion displacement induces the doublet of polaron 
bound states even if the ion vibrates in a harmonic potential, which may be called as the polaron doublet (PD). 
It was shown that the resultant effective potential for the ion displacement behaves like a double-well potential. Therefore, 
the model proposed by Yu and Anderson can be mapped to the TLS and is expected to show the nonmagnetic Kondo effect, which 
will be referred to as Yu-Anderson-type Kondo (YAK) effect.

In this paper, we will study the effects of anharmonicity of a cage potential and discuss low-energy properties of a magnetic ion 
coupled with spinful conduction electrons. The cage potential is assumed to be of the double-well type, where the two minima are 
located symmetrically around the center. By applying numerical renormalization group method\cite{Wilson,krishna} to the present 
model, we find that in the noninteracting case, two types of nonmagnetic Kondo effect mentioned above are realized and the 
low-energy properties make continuous crossover between them when the shape of the cage potential is changed. Then, the role 
of Coulomb interaction $U$ is investigated in the two distinct potential shapes where typical behaviors of the DWK and YAK 
effects are observed in the noninteracting case. We find that only in the typical YAK region, the $2$-channel Kondo fixed 
point (2ch-K) appears with increasing $U$. This behavior is similar to the harmonic potential case. A notable point is that 
the YAK effect is more stable against the Coulomb interaction than the harmonic case.

\section{Hamiltonian of the System}
When a magnetic ion with mass $M$ vibrates in a cage potential $V_{\text{cage}}(\boldsymbol{Q})$,
the dynamics of the ion is determined from the following Hamiltonian,
\begin{align} 
\mathit{H}_{\text{ion}}
&=-\frac{\hbar^{2}}{2M}(\boldsymbol{\nabla}_{\boldsymbol{Q}})^{2}
                                 +V_{\text{cage}}(\boldsymbol{Q}).\label{Ham_ion_general}
\end{align}
In the same way as the previous studies\cite{yashiki1,yashiki2}, we consider the situation that
the ion vibrates in a one-dimensional potential. To investigate effects of anharmonicity 
of the cage potential, we expand $V_{\text{cage}}(Q)$ within the fourth-order of the ion displacement 
$Q$,
\begin{align}
V_{\text{cage}}(Q)=\frac{k_{2}}{2}Q^{2}+\frac{k_{4}}{4}Q^{4},
\end{align}
where the cage potential is assumed to be an even function of $Q$ and $k_{2}$ and $k_{4}(>0)$ 
are the expansion parameters.

The shape of the cage potential is classified into two distinct types; single-well type for $k_{2}>0$ and double-well 
(DW) type for $k_{2}<0$. In this paper, we focus on the DW case. 
By introducing the dimensionless ion displacement $q(\equiv\alpha Q)$, 
the Hamiltonian $\mathit{H}_{\text{ion}}$ is transformed into
\begin{align}
\frac{\overline{\omega}}{2}\biggl\{
-\nabla_{q}^{2}+\Delta_{\text{pot}}\biggl[-2\biggl(\frac{q}{q_{0}}\biggr)^2
               +\biggl(\frac{q}{q_{0}}\biggr)^4\biggr]\biggr\},\label{ham_ion_anharmonic}
\end{align}
where $\overline{\omega}$, $q_{0}$ and $\Delta_{\text{pot}}$ are defined by
\begin{align}
\overline{\omega} = \frac{\alpha^2}{M},\ \ 
q_{0} = \alpha\sqrt{-\frac{k_{2}}{k_{4}}},\ \ 
\Delta_{\text{pot}} = \frac{M(k_{2})^{2}}{2\alpha^{2}k_{4}}.
\end{align}
Two potential minima are symmetrically located at $\pm q_{0}$ in the DW potential and the height of
potential barrier between them is $\Delta_{\text{pot}}$.

The ionic eigenstates $|m\bigl>$ and eigenvalues $E^{(m)}_{\text{ion}}$ are obtained from 
the $\mathrm{Schr\ddot{o}dinger}$ equation of the ion system. We use the convention that $m$ is numbered from zero
in ascending order of the eigenvalues. Because of the relation $V_{\text{cage}}(Q)=V_{\text{cage}}(-Q)$, 
the eigenstates of the ion $|m\bigl>$ are characterized by the inversion, $q\rightarrow-q$. 
The operation of the inversion $\mathcal{P}_{\text{ion}}$ is described by 
\begin{align}
\mathcal{P}_{\text{ion}}|m\bigl> &=  |m\bigl>,\ \ \ \ (m=\text{even})\\
\mathcal{P}_{\text{ion}}|m\bigl> &= -|m\bigl>.\ \ (m=\text{odd})
\end{align}
Therefore, we define the parity of $|m\bigl>$ as $P_{\text{ion}}=0$ $(1)$ depending on 
$m=\text{even}$ $(\text{odd})$.

We have already discussed the derivation of a generalized impurity Anderson model for a magnetic ion
vibrating in a cage potential\cite{yashiki1,yashiki2}. Within the first-order of the ion displacement,
the full Hamiltonian for a magnetic ion with an $s$-wave impurity electron orbital is written by
\begin{align}
    \mathit{H}
=&  \mathit{H}_{\text{c}}+\mathit{H}_{\text{hyb}}+\mathit{H}_{\text{local}},\label{full_Ham_1}\\
    \mathit{H}_{\text{c}}
=&  \sum_{k\sigma} \varepsilon(k)\bigl\{c^{\dagger}_{0\sigma}(k)c_{0\sigma}(k)
                                       +c^{\dagger}_{1\sigma}(k)c_{1\sigma}(k)\bigr\},\\
    \mathit{H}_{\text{hyb}} 
=&  \sum_{k\sigma}\bigl\{V_{0}c^{\dagger}_{0\sigma}(k)f_{\sigma}
                        +V_{1}c^{\dagger}_{1\sigma}(k)f_{\sigma}q+\text{h.c.}\bigr\},\\
    \mathit{H}_{\text{local}}
=&  \mathit{H}_{\text{ion}}
  + \varepsilon_{f}\sum_{\sigma}f^{\dagger}_{\sigma}f_{\sigma} + 
    Uf^{\dagger}_{\uparrow}f_{\uparrow}f^{\dagger}_{\downarrow}f_{\downarrow},\label{full_Ham_2}
\end{align}
where $c_{0\sigma}(k)$ $(c_{1\sigma}(k))$ is an annihilation operator of the $s$-wave ($p$-wave) 
conduction electrons and $\varepsilon(k)$ is assumed to be the dispersion relation with no angle dependence in 
the momentum space. For the localized impurity orbital, the annihilation (creation) operator is 
expressed by $f_{\sigma}$ $(f^{\dagger}_{\sigma})$ and $\epsilon_{f}$ is its energy and $U$ the 
Coulomb interaction. Note that the definition of $V_{1}$ in the present paper is different from Refs. \ref{yashiki1} 
and \ref{yashiki2} by $\sqrt{2}$ and identical to $\overline{V_{1}}$ in Ref. \ref{yashiki2}.

\section{Numerical Renormalization Group Approach}
To investigate the effects of the anharmonicity of the cage potential, we apply the numerical renormalization 
group (NRG) method\cite{Wilson,krishna} to the present model. The merit of the NRG algorithm is that it
enables us to calculate the low-energy spectra and various physical quantities at finite temperatures 
with high accuracy in a controlled way. The key idea of the NRG is to discretize the continuous conduction bands 
in the logarithmic energy scales characterized by $\varLambda$. The discretized Hamiltonian is composed of the 
impurity site and the two Wilson chains. The two chains correspond to the $s$-wave and $p$-wave conduction bands, 
which are coupled with the impurity site through the usual hybridization and the ion displacement-assisted 
one, respectively. The discretized Hamiltonian is written by
\begin{align}
&\mathit{H_{N}}=\varLambda^{\frac{N-1}{2}}\notag\\
 \times&\Biggl\{\ \sum^{N-1}_{n=0,\sigma}\varLambda^{-\frac{n}{2}}\xi_{n}
 \bigl[s^{\dagger}_{n,\sigma}s_{n+1,\sigma}
      +p^{\dagger}_{n,\sigma}p_{n+1,\sigma}+\text{h.c.}\bigr]\notag\\
&\ +\sum_{\sigma}\bigl[ \widetilde{V_{0}}s^{\dagger}_{0,\sigma}f_{\sigma}
                       +\widetilde{V_{1}}p^{\dagger}_{0,\sigma}f_{\sigma}\sum_{m,n}q_{m,n}X_{m,n}
                       +\text{h.c.}\bigr]\notag\\
&\ +\frac{\widetilde{U}}{2}\biggl(\sum_{\sigma}f^{\dagger}_{\sigma}f_{\sigma}-1\biggr)^{2}
   +\biggl(\widetilde{\varepsilon_{f}}+\frac{\widetilde{U}}{2}\biggr)
           \sum_{\sigma}f^{\dagger}_{\sigma}f_{\sigma}\notag\\
&\ +\sum_{m}\widetilde{E}^{(m)}_{\text{ion}}X_{m,m}\Biggr\},\label{Wilson_chain}
\end{align}
where the $n$-th hopping matrix element $\xi_{n}$ is given by
\begin{align}
\xi_{n}=\frac{1-\varLambda^{-n-1}}{\sqrt{1-\varLambda^{-2n-1}}\sqrt{1-\varLambda^{-2n-3}}},
\end{align}
and the matrix element $q_{m,n}$ and the Hubbard operator $X_{m,n}$ are defined by 
\begin{align}
q_{m,n} & \equiv \bigl<m|q|n\bigl>,\\
X_{m,n} & \equiv |m\bigl>\otimes\bigl<n|.
\end{align}
Here, the density of states of the conduction bands, $\rho$, is assumed to be a constant, $1/2D$, 
with the band width of $2D$. All the parameters with tilde are multiplied by the constant factor 
$2/\{D(1+\varLambda^{-1})\}$.

We comment on the symmetries of the Hamiltonian $(\ref{Wilson_chain})$. In the previous 
studies\cite{yashiki1,yashiki2}, three conserved quantum numbers are used; the total electron number 
$N^{\text{tot}}$, the $z$ component of total spin $S^{\text{tot}}_{z}$ and the total parity $P$.
The last one is defined by the sum of the number of $p$-wave conduction electrons $N_{p}$ and 
the number of harmonic phonons $N_{\text{ph}}$, $P=N_{p}+N_{\text{ph}}\equiv0$ or $1$ (mod $2$).
For the anharmonic potential case, obviously, we can use $P_{\text{ion}}$ instead of $N_{\text{ph}}$. This 
straightforward replacement is based on the fact that the full Hamiltonian $(\ref{full_Ham_1})$-$(\ref{full_Ham_2})$ 
is invariant under the inversion of the coordinate system.

The Hamiltonian (\ref{Wilson_chain}) is block-diagonalized and each block is characterized by the 
set of quantum numbers, $N^{\text{tot}}$, $S^{\text{tot}}_{z}$ and $P$. We treat the symmetric case, 
$2\varepsilon_{f}+U=0$, and set various parameters for NRG calculations as follows; $\varLambda=3.0$, 
band width $D=1.0$ and $M=15000$ states kept at each NRG step. For the calculations of the ion eigenstates, 
there are two important parameters, the cutoff number for the states kept concerning the ionic oscillations 
and that of the bases spanned by Hermite polynomials. We use sufficiently big cutoff numbers which depend on 
the potential shape controlled by $q_{0}$ and $\Delta_{\text{pot}}$.

As a result of the NRG calculations, we find that there are three types of the low-energy fixed points 
which are identified by analyzing the energy spectra\cite{yashiki1,yashiki2}. They are classified as the $s$-type, 
the $2$-channel Kondo ($2$-chK) type and the $p$-type fixed points. Detailed discussions about the nature of these 
fixed points have been reported in Ref. \ref{yashiki2}.

\section{Crossover Behaviors among Three Different Regimes}
\begin{figure}
   \begin{center}
      \begin{tabular}{cc}
         \resizebox{40.5mm}{!}{\includegraphics{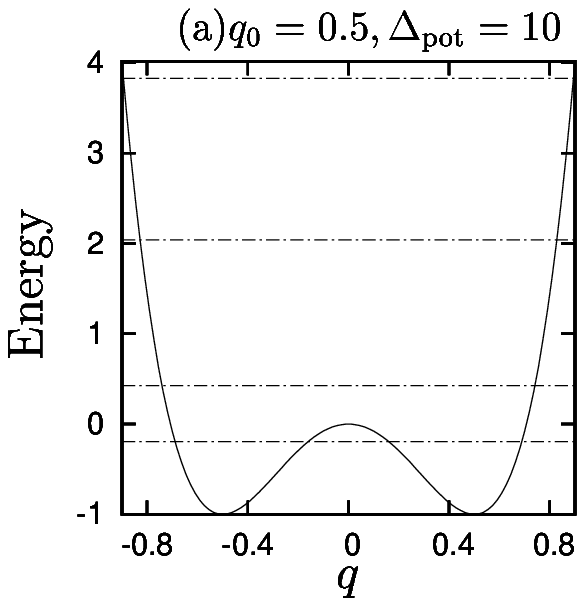}}
         \resizebox{40.5mm}{!}{\includegraphics{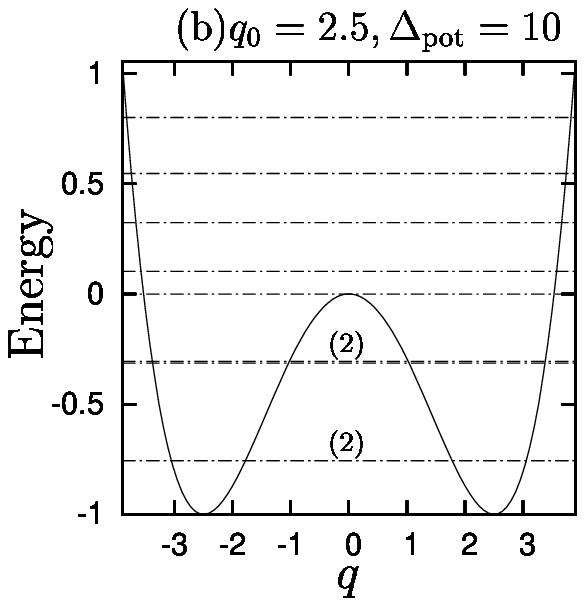}}\\
         \resizebox{75mm}{!}{\includegraphics{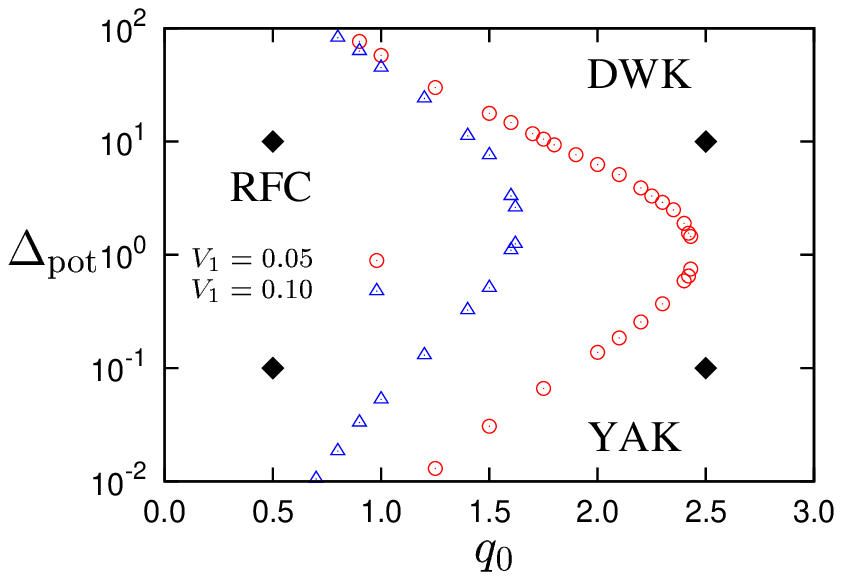}}\\
         \resizebox{40.5mm}{!}{\includegraphics{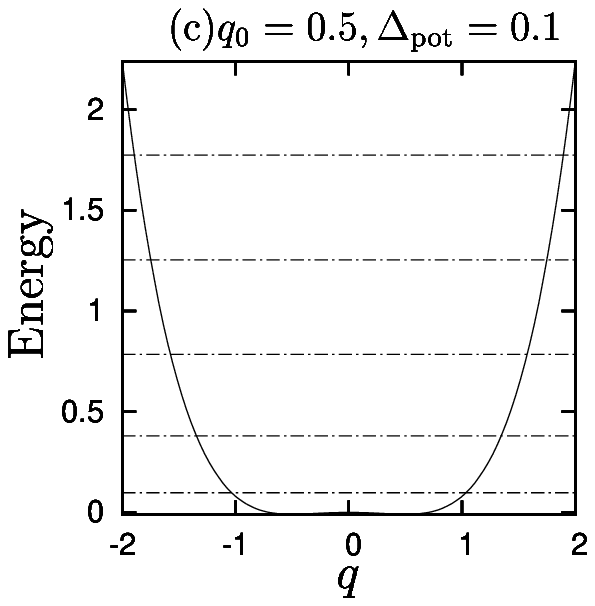}}
         \resizebox{40.5mm}{!}{\includegraphics{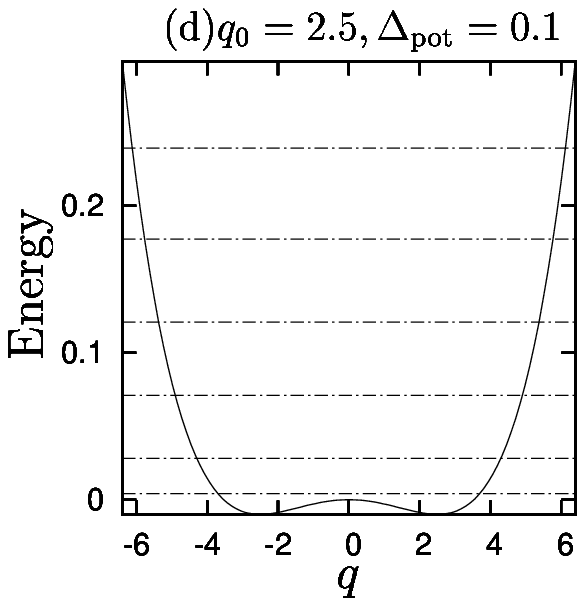}}
      \end{tabular}
   \end{center}
   \caption{(Color online) Phase diagram in the parameter space of $q_{0}$ and $\Delta_{\text{pot}}$
             for $V_{0}=0.2$, $\overline{\omega}=0.2$ and $U=0$ (middle) and the shapes of the anharmonic 
             potential with the low-energy levels represented by the dashed lines at (a) $(q_{0},$ 
             $\Delta_{\text{pot}})=(0.5,10)$, (b) $(2.5,$ $10)$, (c) $(0.5,$ $0.1)$, and (d) $(2.5,$ $0.1)$. These 
             four sets of parameters are represented by the diamonds 
             in the phase diagram. Note that the graphs (a)-(d) are drawn by different scales 
             with regard to the $q$- and $\Delta_{\text{pot}}$-axes. In (b), the dashed lines with the symbol $(2)$ 
             represent almost degenerate energy levels.}\label{phase_diagram}
\end{figure}

First, we discuss the results of the NRG calculations in the noninteracting case $(U=0)$ with $V_{0}=0.2$ 
and $\overline{\omega}=0.2$ fixed. The middle panel in Fig. \ref{phase_diagram} shows the phase diagram in the parameter 
space of $q_{0}$ and $\Delta_{\text{pot}}$. There are two curves represented by the circles and triangles. 
In the left part of the circle (triangle) line, no plateau of $S_{\text{imp}}$ at $k_{\text{B}}\log2$ is seen for 
$V_{1}=0.05$ $(0.10)$. The low-energy fixed point is always of the $s$-type in the entire phase diagram. 
Therefore, the two boundaries are crossover lines.

Around the phase diagram, we show four graphs of the shape of the anharmonic potential with the low-energy levels 
represented by the dashed lines for (a) $(q_{0},$ $\Delta_{\text{pot}})=(0.5,$ $10)$ in the upper left, (b) $(2.5,$ $10)$ 
in the upper right, (c) $(0.5,$ $0.1)$ in the lower left, and (d) $(2.5,$ $0.1)$ in the lower right, respectively. 
The diamonds in the phase diagram correspond to these sets of parameters. 
\begin{figure}[!t]
   \begin{center}
      \begin{tabular}{cc}
         \resizebox{40.5mm}{!}{\includegraphics{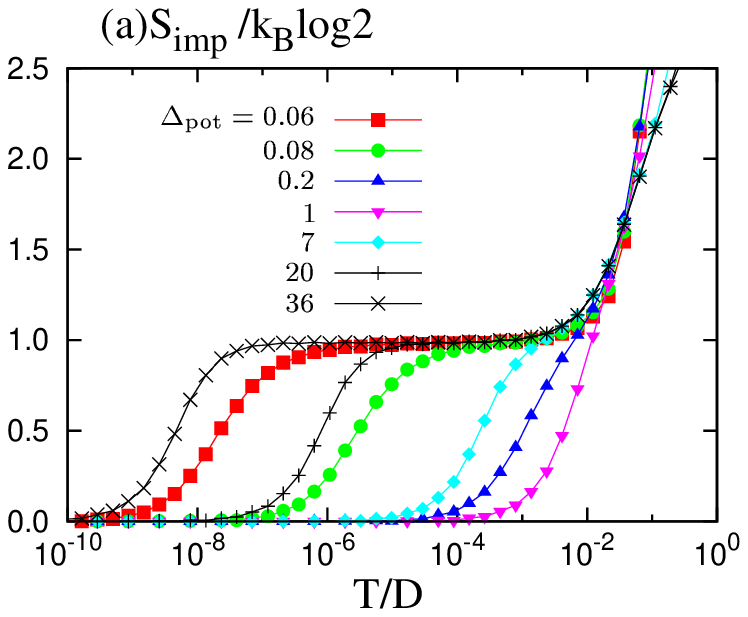}}\\
         \resizebox{40.5mm}{!}{\includegraphics{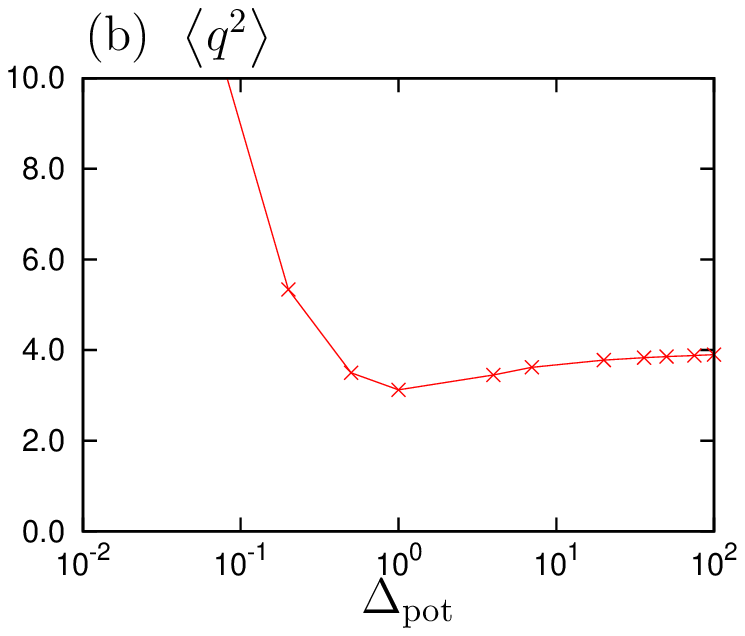}}
         \resizebox{40.5mm}{!}{\includegraphics{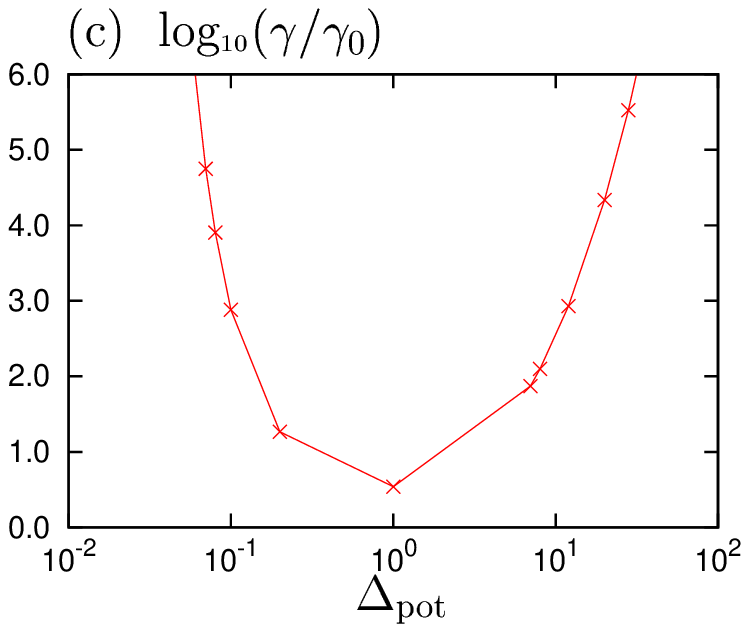}}
      \end{tabular}
   \end{center}
   \caption{(Color online) Temperature dependence of the impurity entropy $S_{\text{imp}}$ for various $\Delta_{\text{pot}}$ values 
             with $V_{1}=0.05$, $U=0$ and $q_{0}=2.0$ in the graph (a). The graph (b) shows the thermal average of the square of the ion 
             displacement $\bigl<q^2\bigr>$ at low temperatures. In the graph (c), $\Delta_{\text{pot}}$ dependence of 
             $\log_{10}(\gamma/\gamma_{0})$ is 
             shown, where $\gamma$ is the specific heat coefficient calculated from the impurity entropy $S_{\text{imp}}$ and 
             $\gamma_{0}$ is that for $V_{1}=U=0$ case.}\label{S_imp_and_S_ion}
\end{figure}

In Fig. \ref{S_imp_and_S_ion}(a), temperature dependence of $S_{\text{imp}}$ is shown for various $\Delta_{\text{pot}}$
with $V_{1}=0.05$, $U=0$ and $q_{0}=2.0$ fixed. We may classify the parameter space into three characteristic regions depending on 
$\Delta_{\text{pot}}$ by observing temperature dependence of $S_{\text{imp}}$. In Fig. \ref{S_imp_and_S_ion}(b), $\Delta_{\text{pot}}$ 
dependence of the thermal average of the square of the ion displacement $\bigl<q^2\bigr>$ at low temperatures is shown.

We note first that for $\Delta_{\text{pot}}<1$, the impurity entropy $S_{\text{imp}}$ is released from the plateau at $k_{\text{B}}\log2$ 
with the same temperature dependence as the usual Kondo effect. In this region, excitation energies of the ion do not show the character 
of the two local minima because the potential barrier between them, $\Delta_{\text{pot}}$, is low [see Figs. \ref{phase_diagram}(c) 
and \ref{phase_diagram}(d)]. In fact, the energy gap between $E^{(0)}_{\text{ion}}$ and $E^{(1)}_{\text{ion}}$ is comparable to the 
difference between $E^{(1)}_{\text{ion}}$ and $E^{(2)}_{\text{ion}}$. Since the DW potential is shallow, the energy spectrum is not 
qualitatively different from the single well potential. Figure \ref{S_imp_and_S_ion}(b) shows that the origin of the plateau is attributed 
not to the two potential minima induced by the anharmonicity of the potential but to the PD induced by combination of many excited states 
of the ion since the $\bigl<q^2\bigr>$ values are significantly enhanced over $4(=q^{2}_{0})$. Note that the distance from the center of the potential 
to one of the local minima is $2(=q_{0})$. From these results, it is reasonable to conclude that physical properties of the impurity is described 
by the YAK effect. The role of the anharmonicity is to make the effect of the electron-vibration coupling stronger and the YAK effect 
is observed even for relatively small coupling constant $V_{1}$.

When $\Delta_{\text{pot}}$ is close to $1$, no plateau appears in $S_{\text{imp}}$. In this parameter region, the present 
system shows the characteristic behaviors that the $s$-wave Wilson chain including the $f$-orbital is perturbed by the $p$-wave Wilson chain 
through the electron-vibration coupling. From the analogy to the harmonic case, we call this physical situation as a renormalized 
Fermi chain (RFC) regime. 

Lastly, when $\Delta_{\text{pot}}$ is larger than $7$, the energy gap between $E^{(0)}_{\text{ion}}$ and $E^{(1)}_{\text{ion}}$, 
$\Delta E_{\text{ion}}$, becomes small, which is prominent in Fig. \ref{phase_diagram}(b). Figure \ref{S_imp_and_S_ion}(a) shows that 
temperature dependence of $S_{\text{imp}}$ for $\Delta_{\text{pot}}\geq7$ is identical to that of the usual spin Kondo effect under a weak 
magnetic field. With increasing $\Delta_{\text{pot}}$, the origin of the plateau of $S_{\text{imp}}$ at $k_{\text{B}}\log2$ changes from 
the PD to the double-well potential minima. Correspondingly, Figure \ref{S_imp_and_S_ion}(b) shows that for $\Delta_{\text{pot}}\gg1$, 
$\bigl<q^2\bigr>$ converges to $4$. Therefore, the upper right region of the parameter space may be characterized by the double-well-type 
Kondo (DWK) regime.

Actually, we can confirm realization of the DWK effect from the specific heat. $\Delta_{\text{pot}}$ dependence of 
$\log_{10}(\gamma/\gamma_{0})$ with $q_{0}=2.0$ is shown in Fig. \ref{S_imp_and_S_ion}(c), where $\gamma$ is 
the specific heat coefficient and $\gamma_{0}$ is that for $V_{1}=U=0$ case. In this study, 
$\gamma$ is calculated from central differences of the averaged impurity entropy $\overline{S}_{\text{imp}}(T_{i})$ at the temperature 
$T_{i}$ corresponding to the $i$-th NRG step,
\begin{align}
\gamma(\overline{T_{i}})=\frac{\overline{S}_{\text{imp}}(T_{i})-\overline{S}_{\text{imp}}(T_{i+1})}{\log{\sqrt{\varLambda}}},
\end{align}
where $\overline{S}_{\text{imp}}(T_{i})$ is defined by 
$\frac{1}{2}S_{\text{imp}}(T_{i})+\frac{1}{4}(S_{\text{imp}}(T_{i-1})+S_{\text{imp}}(T_{i+1}))$. $\overline{T_{i}}$ is the averaged 
temperature on a logarithmic scale, $\sqrt[4]{\varLambda}T_{i}$. The minimum of $\gamma$ is located in the RFC regime, where $\gamma$ is 
weakly enhanced by the electron-vibration coupling in comparison with the noninteracting Fermi liquid. The figure reveals that the picture 
of the local Fermi liquid is valid for the entire parameter region. The electron-vibration coupling leads to an effectively heavy local Fermi 
liquid state at low temperatures by the mechanism of the YAK for small $\Delta_{\text{pot}}$ or the DWK for large $\Delta_{\text{pot}}$.

Turning back to the phase diagram of Fig. \ref{phase_diagram}, we comment on the $V_{1}$ dependence of the boundaries determined from 
$S_{\text{imp}}$. For $q_{0}\lesssim1.0$ and $\Delta_{\text{pot}}>1$, the boundaries determined from $S_{\text{imp}}$ have only 
a weak $V_{1}$ dependence. This is consistent with the fact that the effect of the electron-vibration coupling is effectively 
weakened when the space where the ion can move becomes too small. Strong $V_{1}$ dependence appears for $q_{0}\gtrsim1.0$. The parameter 
region where there is no plateau at $k_{\text{B}}\log2$ is expanded with decreasing $V_{1}$.

With regard to the double-well potential case, what type of the low-energy fixed point is realized is a subtle question. 
Let us start the discussion from the mapping of the present Hamiltonian $(\ref{full_Ham_1})$-$(\ref{full_Ham_2})$ to the TLS. 
Neglecting higher ion excited states above the first 
one, we introduce the pseudo-spin to describe the almost doubly degenerate ionic eigenstates located at the two potential minima. 
Concerning the ion state, the ground state with the even parity is chosen as the eigenstate of the $z$-component of the pseudo-spin,
$\tau^{i}_{z}|0\bigl>=-|0\bigl>$. In the same way, $\tau^{i}_{z}|1\bigl>=|1\bigl>$. Similarly, we can define the pseudo-spin for the
conduction electrons. For this purpose, we define $a_{0\sigma}=f_{\sigma}$ and $a_{1\sigma}=\sum_{k}c_{1\sigma}(k)$.

Under this approximation, the operator of the ion displacement $q$ corresponds to the $x$ component of the Pauli matrix for the ion
$\tau^{\text{i}}_{x}$. Then, the Hamiltonian $(\ref{full_Ham_1})$-$(\ref{full_Ham_2})$ can be rewritten as
\begin{align}
 \mathit{H}=
& V_{1}\overline{q}\sum_{\sigma}\sum_{\alpha,\beta=0,1}a^{\dagger}_{\alpha\sigma}[\tau^{\text{e}}_{x}]_{\alpha\beta}
  a_{\beta\sigma}\cdot\tau^{\text{\text{\text{i}}}}_{x}
 +\frac{\Delta E_{\text{ion}}}{2}\tau^{\text{i}}_{z}\notag\\
&+\mathit{H}_{\text{c}}+V_{0}\sum_{k\sigma}\bigl\{c^{\dagger}_{0\sigma}(k)a_{0\sigma}+\text{h.c.}\bigr\},\label{TLS_ham}
\end{align}
where $\tau^{\text{e}}_{x}$ is the $x$ component of the Pauli matrix for the conduction electrons and $\overline{q}$ is the matrix 
element of $q$ between $|0\bigl>$ and $|1\bigl>$. The second term represents the energy difference between $|0\bigl>$ and $|1\bigl>$, 
which originates from the tunneling process between the two potential minima.

To begin with, we discuss the original TLS proposed by Vlad$\acute{\text{a}}$r and Zawadowski\cite{TLM1,TLM2,TLM3} for the spinless 
fermion case. By considering scattering processes of the second-order of the ion displacement, they showed that the TLS can be mapped 
to the anisotropic Kondo model under the pseudo-field given by $\Delta E_{\text{ion}}$. Because of the existence of the impurity 
$f$-orbital, the present model does not have the channel symmetry between the $s$-wave and $p$-wave conduction electrons\cite{yashiki2} 
unlike the TLS. This asymmetry in the channel space is sufficient to realize the Kondo effect for the present model even in the absence 
of processes of the second-order of displacement. Similar discussion is presented in the previous studies on the Yu-Anderson-type Kondo 
effect\cite{yashiki1,yashiki2}.

Now, we go to the spinful case. Since the real spin can be used as the channel index for the conduction electrons, the non-Fermi 
liquid behaviors of the $2$-chK may be expected for the spinful fermion case. It is known that the anisotropy of the (pseudo-)spin space is 
irrelevant to the $2$-chK fixed point\cite{Affleck}. In the present NRG results, we have not seen any indication of the $2$-channel 
Kondo effect for $U=0$. We may argue that there are three reasons for the absence of the $2$-chK fixed point. Firstly, 
$\Delta E_{\text{ion}}$ works as the transverse pseudo-field acting on the pseudo-spin. The non-Fermi liquid behaviors may be 
quenched by this external pseudo-field. Secondly, the eigenstates of the ion above the first excitation can not be neglected 
because the ion excitation energies are almost always lower than the band width $D$ and plays the role of energy 
cutoff\cite{Aleiner1,Aleiner2}. Aleiner $et$ $al.$ showed that when higher ion excited states more than the first one are included 
into the TLS for the spinful fermions, the $2$-chK temperature is always lower than $\Delta E_{\text{ion}}$. The Third reason is the 
channel asymmetry which is relevant to the $2$-chK fixed point\cite{Affleck}.
\begin{figure}[t]
   \begin{center}
      \begin{tabular}{cc}
         \resizebox{40.5mm}{!}{\includegraphics{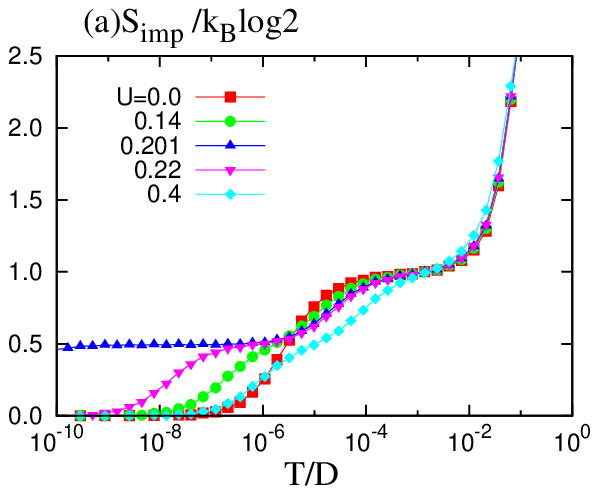}}
         \resizebox{40.5mm}{!}{\includegraphics{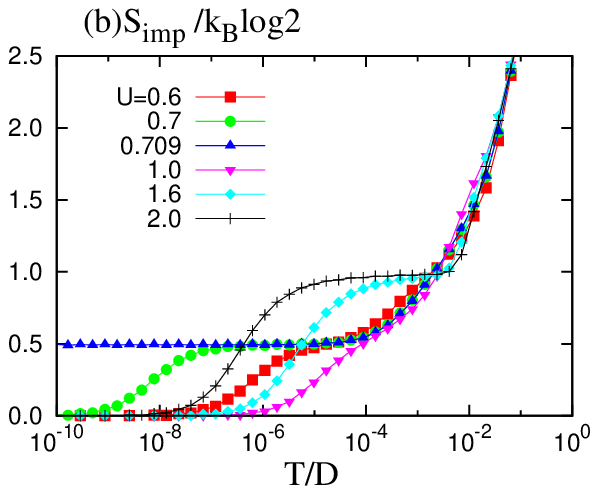}}\\
         \resizebox{40.5mm}{!}{\includegraphics{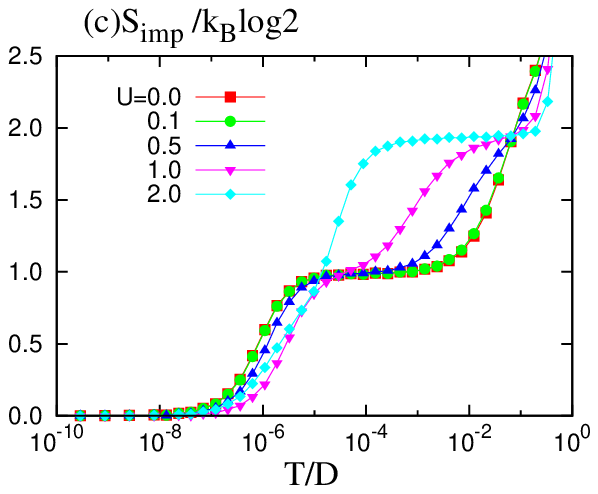}}
      \end{tabular}
   \end{center}
   \caption{(Color online) Temperature dependence of the impurity entropy $S_{\text{imp}}$ for $0.0<U<0.4$ in the upper left 
             graph (a) and $0.6<U<2.0$ in the upper right one (b) with $q_{0}=2.0$ and $\Delta_{\text{pot}}=0.08$. That for 
             $\Delta_{\text{pot}}=20$ is shown in the lower graph (c).}\label{U_dependence}
\end{figure}

Lastly, we investigate effects of the Coulomb interaction $U$ on physical properties of the impurity site. From the NRG calculations, 
we find that the responses against $U$ depend on the potential shapes of the cage. Figure \ref{U_dependence} shows temperature 
dependence of the impurity entropy $S_{\text{imp}}$ for various $U$ values with $V_{1}=0.05$ and $q_{0}=2.0$. In the upper two graphs 
(a) and (b), $\Delta_{\text{pot}}=0.08$, and in the lower graph (c) $\Delta_{\text{pot}}=20$.

For $\Delta_{\text{pot}}=0.08$, we observe the $2$-channel Kondo fixed points at $U=0.201$ and $0.709$ 
from the analyses of the low-energy spectra. The type of the energy spectra changes from the $s$-type to the $p$-type at $U=0.201$ 
and return back to the $s$-type at $0.709$ with increasing $U$. Such $U$ dependence of the low-energy fixed points is the same as 
that for the harmonic case\cite{yashiki1,yashiki2}. In comparison with the harmonic case, there is one major quantitative 
difference about the lower critical $U$ of the $2$-chK fixed point. For the harmonic case, the lower critical $U$ is roughly 
estimated to be the same order as the YAK temperature for the noninteracting case, see section 4.7 in Ref. \ref{yashiki2}.
This picture is not valid for the anharmonic case as shown in Fig. \ref{U_dependence}(a). It seems that the anharmonicity of 
the cage potential allows the YAK effect to be robust against $U$, in particular for the shallow potential case.

Completely different $U$ dependence is observed for the case $\Delta_{\text{pot}}=20$, as shown in Fig. \ref{U_dependence}(c). 
When $U$ increases, a new plateau of $S_{\text{imp}}$ at $2k_{\text{B}}\log 2$ appears. The value $2k_{\text{B}}\log 2$ indicates 
that physical state of the impurity site is expressed by the direct product of the local magnetic moment for the electron part and 
the two potential minima for the ion part. From the low-energy spectra, the fixed point is always of the $s$-type regardless of $U$. 
Temperature dependence of the entropy release from the plateau at $2k_{\text{B}}\log 2$ is controlled by the two energy scales 
corresponding to the $s$-channel Kondo temperature and $\Delta E_{\text{ion}}$.

These two different types of responses against $U$ correspond to the results of the noninteracting case. In the YAK regime, 
the physical state of the impurity is considered to be the PD which consists of the spin singlet of the electron part and 
the linear combination of many excited states of the ion. By sharing the ion displacement by up and down spin electrons, 
there is an effectively attractive interaction, which leads to the energy lowering for the singlet channel. Introduction of 
repulsive $U$ changes the situation, and at the critical $U$, the singlet and triplet channels become degenerate. In this way, 
when $U$ increases, the $2$-chK fixed point is realized. 

On the other hand, in the DWK regime, the ion moves back and forth between the two potential minima. The ionic eigenstates are 
effectively written by only the almost degenerate states and the higher ion excited states are not active. The real spin degrees 
of freedom define two independent screening channels for the pseudo-spin. However, in the noninteracting case, 
the $2$-chK fixed point is suppressed by the three reasons mentioned previously. When $U$ increases, the up spin and the down spin 
electrons are correlated. Therefore, under the finite $U$, we can not use the real spin degrees of freedom as the independent channel 
index for the conduction electrons, and the $2$-chK fixed point is not observed.

\section{Conclusions}
We have studied the generalized Anderson model constructed for a magnetic ion vibrating in the double-well potential.  By using the 
NRG method, we find that, for $U=0$, the low-energy fixed point is always of the $s$-type and that the low-energy physics are described 
by the local Fermi liquid theory. The NRG results on $S_{\text{imp}}$ and $\bigl<q^2\bigr>$ show that the parameter space can be 
classified into three different regimes: the YAK, DWK and RFC regimes. However, there is no sharp boundary between the three regimes 
and only smooth crossover behaviors are observed. 

One interesting question arises concerning the DWK regime. Since the effective Hamiltonian obtained by restricting to the two lowest 
vibrating states is represented by the TLS with the real spin as an additional channel index, one may expect realization of the non-Fermi 
liquid behaviors associated with the $2$-chK fixed point. However, the present NRG results do not show any indication of the $2$-chK 
fixed point. We conclude that the renormalization flow to the $2$-chK fixed point is terminated by the following three factors: the 
transverse pseudo-field $\Delta E_{\text{ion}}$, the higher ion excited states and the channel asymmetry inherent to the present model.  
 
Subsequently, we have considered the effect of Coulomb interaction $U$ in the YAK and DWK regimes. In the YAK regime, by the same 
mechanism as the harmonic potential case, the $2$-chK fixed point appears twice with increasing $U$. One interesting result is that 
the lower critical $U$ of the $2$-chK is considerably enhanced in comparison with the harmonic case, even if the coupling constant 
$V_{1}$ is not so large. We conclude that the vibration of a magnetic ion in an oversized cage structure provides us an opportunity 
to realize the $2$-chK effect.
On the other hand, the fixed point remains always of the $s$-type when $U$ increases in the DWK regime. The present 
results indicate that the anharmonicity of the cage potential induces effectively strong electron-vibration coupling, especially in the 
shallow potential case. Therefore, we may conclude that the anharmonicity of the cage potential is favorable to the realization of the 
nonmagnetic Kondo effect, which is one of the promising candidates for the magnetically robust heavy Fermion behavior observed in the 
skutterudite compound\cite{Sm_Os4_Sb12_2}.

\section*{Acknowledgements}
The authors would like to thank Shunsuke Kirino for his supports concerning the NRG calculations, and Kazumasa Hattori and 
Takashi Hotta for useful discussions.
This work is supported by Grant-in-Aid on Innovative Areas "Heavy Electrons" (No.$20100208$) and also by Scientific Research (C) 
(No.$20540347$). S.Y. acknowledges support from Global COE Program "the Physical Sciences Frontier", MEXT.

\end{document}